\newcommand{\be}[0]{\begin{equation}}
\newcommand{\ee}[0]{\end{equation}}
\newcommand{\ba}[0]{\begin{eqnarray}}
\newcommand{\ea}[0]{\end{eqnarray}}
\documentstyle[amssymb,11pt,epsfig,epsf]{article}

\input epsf

\begin{document}
\title{\bf{Extracting the QCD ${\Lambda _{\overline {MS} }}$ parameter in Drell-Yan Process using Collins-Sopper-Sterman approach
 }}
\author{R. Taghavi\thanks{%
r.taghavi@stu.yazd.ac.ir} $^{(a)}$, A. Mirjalili\thanks{%
a.mirjalili@yazd.ac.ir} $^{(a)}$  \and \vspace{0.15in}\vspace{0.07in} \\
$^{(a)}$ Physics Department, Yazd University, Yazd, Iran \\
}
\maketitle
\begin{abstract}
In this work we directly fit the QCD dimensional transmutation parameter, ${\Lambda _{\overline {MS} }}$, to experimental data of Drell-Yan (DY) observables. For this purpose  we first obtain the evolution of transverse momentum dependent parton distribution functions (TMDPDFs) up to the NNLO approximation based on Collins-Sopper-Sterman formalism. As is expecting the  TMDPDFs are appearing at larger values of transvers momentum by increasing the energy scales and also the order of approximation. Then we calculate the cross section related to the TMDPDFs in the DY process. As a consequence of fitting  to the E288 experimental data at center of mass energy  $\sqrt{s}=23.8$  $GeV$, we obtain  ${\Lambda _{\overline {MS} }} = 249 \pm 7 MeV$ corresponding to the renormalized coupling constant ${\alpha _s}(M_z^2) = 0.119 \pm 0.001$ which is within the acceptable range for this quantity. The results for DY cross section at different   energy  scales are in good agreement within the available data.
\end{abstract}
\section{ Introduction}
To discuss the details of probing  hadronic structure in high energy collisions, the intrinsic transverse momentum carried by partons plays an essential role.
Allowing  the parton distribution functions (PDFs) and  fragmentaion function (FFs) to depend  additionaly on intrinsic transverse momentum rather than  the usual momentum fraction variables will cause the teransverse momentum dependent (TMD) factorization formalism  go beyond the framework of standard factorization \cite{1o,2o}. To describe the processes which are sensitive to intrinsic parton transverse, TMD factorization will be needed.
There are large variety of situations which can be considered as  an eveidnent to appear the usefulness of the TMD concept. Some examples  are  Drell-Yan (DY) process, hadron production in electron-positron annihilation at small transverse momentum and  semi inclusive deep inelastic scattering (SIDIS).
In this article, we are focusing on the DY processes and try to do some phenomenological investigation to extract an improtant parameter in perturbative QCD analysis.

The renormalized strong coupling constant ${\alpha _s}(\mu )$ is a crucial quantity in high energy collisions . Equivalently the Renormalization Group (RG) invariant scale parameter ${\Lambda _{\overline {MS} }}$  is the fundamental QCD scale which is depend on the number of active quark flavour, ${n_f}$, and can be extracted using fit to the different experimental data confronted with theoretical predictions.  Although lots of works has been done to determine this parameter, it is still of interest to be estimated  by other methods and using different available experimental data.  Since the Drell-Yan process is one of the good context to test QCD theory, we choose the available low energy experimental data related to this process to find the ${\Lambda _{\overline {MS} }}$ parameter by makeing use of Collins, Sopper and Sterman(CSS) resummation formalism \cite{1} which is one of the theoretical framework designed to account for QCD effects. The fitting is performed at first four mass bins of 5 $GeV$ to 9 $GeV$ of the E288 experimental data . Since the used energy scale, $Q$, is more than the mass of bottom quark, five active flavour is considered in our analysis. Totally 28 data points are taken into account to extract ${\Lambda _{\overline {MS} }}$ in the fit.

The organization of paper is as following. In Sec.2 we consider the evolution of TMD PDF which is employed to calculate the cross section in DY process by making use of the  Collins-Sopper-Sterman (CSS) approach. We  extend the current results in \cite{3} and obtain the TMD PDF up to the NNLO approximation. In Sec.3 we review the theoretical assumption to calculate the cross section of DY processes in which the Brock-Landry-Nadolsky-Yuan (BLNY) parametrization is used for the non-perturbative part of the calculation. Using fitting process the unknown parameter ${\Lambda _{\overline {MS} }}$ is obtained in Sec.4. Finally we give our conclusion in Sec.5.
\section{TMD evolution}
The evolution of quark-TMD PDF  in the CSS formalism  is given by \cite{3,4}:
\begin{eqnarray}
{{\tilde F}_{f/P}}(x,{b_T};\mu ,{\zeta _F}) = \overbrace {\sum\limits_j {\int_x^1 {\frac{{d\hat x}}{{\hat x}}} } {{\tilde C}_{f/j}}(x/\hat x,{b_ * };{\mu _b}^2,{\mu _b},g({\mu _b})){f_{j/P}}(\hat x,{\mu _b})}^{(a)}\nonumber\\
 \times \overbrace {\exp \{ \ln \frac{{\sqrt {{\zeta _F}} }}{{{\mu _b}}}\tilde K({b_ * };{\mu _b}) + \int_{{\mu _b}}^\mu  {\frac{{d{\mu ^{'}}}}{{{\mu ^{'}}}}} [{\gamma _F}(g({\mu ^{'}});1) - \ln \frac{{\sqrt {{\zeta _F}} }}{{{\mu ^{'}}}}{\gamma _K}(g({\mu ^{'}}))]\} }^{(b)}\nonumber\\ \times \overbrace {\exp \{ {g_{j/P}}(x,{b_T}) + {g_K}({b_T})\ln \frac{{\sqrt {{\zeta _F}} }}{{\sqrt {{\zeta _{F,0}}} }}\} }^{(c)}.\label{f}
\end{eqnarray}
In this equation, the ${f_{j/P}}(\hat x,{\mu _b})$ is the ordinary PDF and   ${\tilde C_{f/j}}(x/\hat x,{b_ * };{\mu _b}^2,{\mu _b},g({\mu _b}))$, $\tilde K({b_ * };{\mu _b})$, ${\gamma _K}(g({\mu ^{'}}))$
and ${\gamma _F}(g({\mu ^{'}});1)$ are functions which  for all ${b_T}$ are  perturbatively calculable. The non-perturbative ${b_T}$ behavior of ${\tilde F_{f/P}}(x,{b_T};\mu ,{\zeta _F})$ and $K(b;{\mu _b})$  are governed  respectively  by the functions ${g_{j/P}}(x,{b_T})$ and ${g_K}({b_T})$ . These functions are scale-independent and universal. The anomalous dimensions of the ${{\tilde F}_{f/P}}(x,{b_T};\mu ,{\zeta _F})$ and $\tilde K({b_ * };{\mu _b})$ are represented   by   the ${\gamma _F}$ and ${\gamma _K}$ respectively. In the small ${b_T} \ll \frac{1}{{{\Lambda _{QCD}}}}$  limit   the first phrase, $(a)$,  in Eq.(\ref{f}) matches the TMD PDF to a collinear treatment . The hard part of this phrase is given by Wilson coefficient function, ${\tilde C_{f/j}}$, where it also contains the standard integrated PDF which is denoting  a collinear factor. The second phrase, $(b)$, of Eq.(\ref{f}) includes exponential functions which are different and  all can perturbatively be calculated . The matching between the small and large ${b_T}$-dependence  is implementing by the third phrase, $(c)$, of this equation. The non-perturbative feature, intrinsic to the proton,  for large-${b_T}$   values  is given by ${g_{j/P}}(x,{b_T})$ function. On the other hand the non-perturbative behavior  of $\tilde K({b_ * };{\mu _b})$ at large values of ${b_T}$ is presented by ${g_K}({b_T})$. However the function ${g_{j/P}}(x,{b_T})$ depends on the external hadron but it can be considered universal. For ${g_K}({b_T})$ function there is different situation in which it is universal and also do not depend on the  external types of the hadrons.

To do the calculations, as it is customary, the choose $\sqrt {{\zeta _F}} = Q$ and $\sqrt {{\zeta _{F,0}}}  = {Q_0}$ are applied. In this work, we first obtain the TMD PDF of up quark  at the NNLO approximation  with x = 0.09 for the  small $Q = \sqrt {2.4}$, medium  $Q = 5$ and  $Q =91.9 $ $GeV$ values. To do this we use the package of $PDFCollinear[parton,x,{b_T}]$ \cite{9} which gives us the collinear part of the TMD PDF for a parton of a specified quark. Since up-quark is concerned in our calculations, the $PDFCollinear[up,x,{b_T}]$ is used for first phrase, $(a)$, of Eq.(\ref{f}). Considering the second phrase, $(b)$, we are able to increase the accuracy of calculation, taking into account  the NNLO contributions of $\tilde K({b_ * };{\mu _b})$ ,${\gamma _F}$  and ${\gamma _K}$ functions.

The one-loop values for $\tilde K({b_ * };{\mu _b})$ ,${\gamma _F}$  and ${\gamma _K}$ functions are as following \cite{4}:
\begin{eqnarray}
\tilde K(\mu ,{b_T}) =  - \frac{{{\alpha _s}{C_F}{L_ \bot }}}{\pi },
\end{eqnarray}
\begin{eqnarray}
{\gamma _F}(\mu ) = \frac{{{\alpha _s}{C_F}}}{\pi }(\frac{3}{2} - \ln (\frac{{{\zeta _F}}}{{{\mu ^2}}})),
\end{eqnarray}
\begin{eqnarray}
{\gamma _K}(\mu ) = 2\frac{{{\alpha _s}{C_F}}}{\pi }\;,
\end{eqnarray}
where:
\begin{eqnarray}
{L_ \bot }(\mu ,b_T) = Log(\frac{{{\mu ^2}{b_T}^2}}{{4{e^{ - 2{\gamma _E}}}}}).
\end{eqnarray}
Considering  Eqs.(2-5) the TMD PDF for up quark at three different scales can be calculated in the NLO approximation.
To reach our main aim to obtain the NNLO approximation in  evolving the TMD PDF, we need to add the NNLO contributions to the Eqs.(2-4) which are available in [5] as:
\begin{eqnarray}
\tilde K(\mu ,{b_T}) =  - \frac{{{\alpha _s}{C_F}{L_ \bot }}}{\pi } + {(\frac{{{\alpha _s}}}{\pi })^2}(\frac{1}{{32}}{\Gamma _0}{\beta _0}{L_ \bot }{(\mu ,{b_T})^2} - \frac{1}{8}(\frac{{{\Gamma _1}}}{2}{L_ \bot }(\mu ,{b_T})) - \frac{1}{8}{d_{20}}),
\end{eqnarray}
\begin{eqnarray}
{\gamma _F}(\mu ;{\zeta _F}/{\mu ^2}) = {\alpha _s}\frac{{{C_F}}}{\pi }(\frac{3}{2} - \ln (\frac{{{\zeta _F}}}{{{\mu ^2}}})) + {(\frac{{{\alpha _s}(\mu )}}{\pi })^2}( - 4{C_F}((\frac{{67}}{9} - \frac{{{\pi ^2}}}{3}){C_A} - \frac{{20}}{9}{T_F}{N_f})(Log[\frac{\zeta }{{{\mu ^2}}}])\nonumber \\
- {C_F}^2( - 3 + 4{\pi ^2} - 48{\varsigma _3}) - {C_F}{C_A}( - \frac{{961}}{{27}} - \frac{{11{\pi ^2}}}{3} + 52{\varsigma _3}) - {C_F}{T_F}{N_f}(\frac{{260}}{{27}} + \frac{{4{\pi ^2}}}{3})),
\end{eqnarray}
\begin{eqnarray}
{\gamma _K}(\mu ) = 2\frac{{{\alpha _s}{C_F}}}{\pi } + 2{(\frac{{{\alpha _s}}}{\pi })^2}({C_F}{C_A}(\frac{{67}}{{36}} - \frac{{{\pi ^2}}}{{12}}) - \frac{5}{{18}}{C_F}{N_f})\;,
\end{eqnarray}
\\where:
\begin{eqnarray}
{\Gamma _0} = 4{C_F},
\end{eqnarray}
\begin{eqnarray}
{\beta _0} = \frac{{11}}{3}{C_A} - \frac{4}{3}{T_F}{N_f},
\end{eqnarray}
\begin{eqnarray}
{\Gamma _1} = 4{C_F}((\frac{{67}}{9} - \frac{{{\pi ^2}}}{3}){C_A} - \frac{{20}}{9}{T_F}{N_f}),
\end{eqnarray}
\begin{eqnarray}
{d_{20}} = {C_F}{C_A}(\frac{{404}}{{27}} - 14{\varsigma _3}) - (\frac{{112}}{{27}}){C_F}{T_F}{N_f}.
\end{eqnarray}
\\By inserting  Eqs.(6-8) in Eq.(1) and doing the required numerical calculations the TMD PDF of up quark can be obtained for the small, medium and large values of $Q = \sqrt {2.4}$ , 5, and 91.19 GeV  with x = 0.09 in ${Q_0} = 1.6$. We depict the related plots run over a range typical for studies of TMD-functions from ${P_T} = 0$ to 6 GeV in Fig.1.

\begin{figure}[htb]
\begin{center}
\vspace{1cm}
\resizebox{0.7\textwidth}{!}{\includegraphics{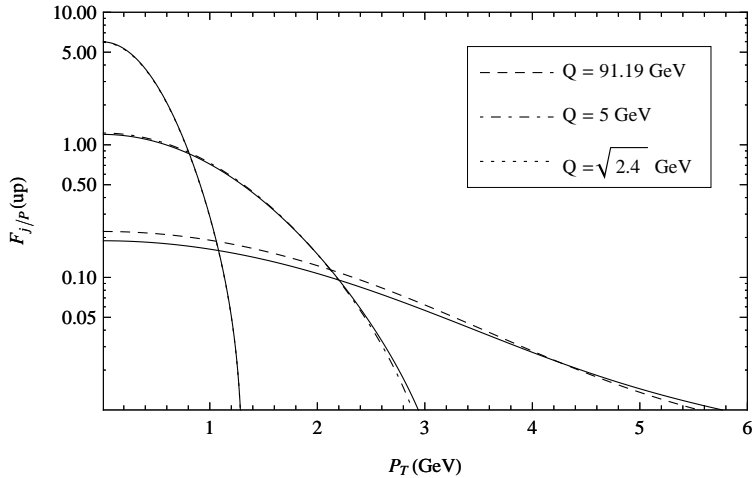}}
\caption{ The  TMDPDF for up quark at $Q = \sqrt {2.4} $ , 5.0 and 91.19 GeV with x = 0.09. The plot shows the result at the NLO (dashed and dash-dotted curves) and NNLO (solid curves) approximations. Solid curves are resulting  from our results with ${b_{\max }} = 0.5\;Ge{V^{ - 1}}$. } \label{JR2009PDF}
\end{center}
\end{figure}

As is expected by increasing the energy scale the effect of TMD PDF at larger values of $P_T$ is growing. By raising the accuracy of calculations and increasing the energy scales, the amount of of TMDPDF at small $P_T$  value is deceasing. In this case the effect of this function at large values of $P_T$ is slightly extending.

\section{Hadronic production and DY processes in CSS formalism}
The differential cross section for the Drell-Yan process in the CSS resummation formalism can be expressed
as \cite{6,6n}:
\begin{eqnarray}
\frac{{d\sigma }}{{d{Q^2}d{Q_T}^2dy}} = \frac{1}{{{{(2\pi )}^2}}}\delta ({Q^2} - {M_V}^2)\int {{d^2}b{e^{i{{\vec Q}_T}.{{\vec b}_T}}}{{\tilde W}_{j\bar k}}(b_T,Q,{x_1},{x_2}) + Y(Q,{Q_T},} {x_1},{x_2}).\label{dy}
\end{eqnarray}
In this equation $Q,{Q_T}$ and $y$ are the invariant mass, transverse momentum and  rapidity of vector boson V respectively in which ${x_1}, {x_2}$ are presented by  ${x_1} = {e^y}Q/s$, ${x_2} = {e^{ - y}}Q/s$ where s is the overall center-of-mass energy of colliding hadrons. In Eq.(\ref{dy}) the two dimensional fourier transform can be converted to one dimensional integral, using the first type of Bessel function in which Eq.(\ref{dy}) can be appeared as \cite{4}:
\begin{eqnarray}
\int {{d^2}{b_T}{e^{i{P_T}.{{\vec b}_T}}}{{\tilde W}_{j\bar k}}({b_T},Q,{x_1},{x_2})}  = 2\pi \int {{b_T}d{b_T}{J_0}({b_T}{P_T}){{\tilde W}_{j\bar k}}({b_T},Q,{x_1},{x_2})}.\label{w}
\end{eqnarray}
The ${{\tilde W}_{j\bar k}}$ term is defined by \cite{5}:
\begin{eqnarray}
{\widetilde W_{j\overline k }}(b_T,Q,{x_1},{x_2}) = {e^{ - S(Q,b_T,{C_1},{C_2})}}\sum\limits_{j,\overline k } {\frac{{{\sigma _0}}}{s}{{\overline P }_{j,{h_1}}}} ({x_1},b_T,\frac{{{C_1}}}{{{C_2}b_T}}){\overline P _{\overline k, {h_2}}}({x_2},b_T,\frac{{{C_1}}}{{{C_2}b_T}}) + (j \leftrightarrow \overline k )\;. \label{w}
\end{eqnarray}
In Eq.(\ref{w}) the indices ${h_1}$ ,${h_2}$ denotes to  the parton number 1 and 2. The $\sum\limits_j {{{\overline P }_{j{h_1}}}} ({x_1},b,\frac{{{C_1}}}{{{C_2}b}})$ corresponds to the part $(a)$ of Eq.(\ref{f}){ where we calculate it, using the package of $PDFCollinear[up,{x_1},{b_T}]$ as we described in Sec.2.} The ${C_1}$ and ${C_2}$ are renormalization constants. To remove some of the logarithmic divergences in ${\widetilde W_{j\overline k }}(b_T,Q,{x_1},{x_2})$, the best choice for these constants are ${C_1} = 2{e^{{\gamma _E}}} \equiv {b_0}$ and ${C_2} = \frac{{{C_1}}}{{{b_0}}} = 1$ \cite{8} . The Sudakov factor S in Eq.(\ref{w}) which is equivalent to the square of the $(b)$ factor in Eq.(\ref{f}) is given by \cite{5}:
\begin{eqnarray}
S(Q,b_T,{C_1},{C_2}) = \int_{{C_1}^2/{b_T^2}}^{{C_2}^2{Q^2}} {\frac{{d{{\overline \mu  }^2}}}{{\overline \mu  }}} [{\rm A}({\alpha _s}(\overline \mu  ),{C_1})\ln (\frac{{{C_2}^2{Q^2}}}{{{{\overline \mu  }^2}}}) + B({\alpha _s}(\overline \mu  ),{C_1},{C_2})]\;. \label{s}
\end{eqnarray}
The A and B functions  in Eq.(\ref{s})   can be found up to NNLO approximation in \cite{6} as:
\begin{eqnarray}
A({\alpha _s}(\bar \mu ),{C_1}) = \sum_{N=1}^2 {A^{(N)}}({C_1}){[\frac{{{\alpha _s}(\bar \mu )}}{\pi }]^N},
\end{eqnarray}
\begin{eqnarray}
B({\alpha _s}(\bar \mu ),{C_1},{C_2}) = \sum_{N=1}^2 {B^{(N)}}({C_1},{C_2}){[\frac{{{\alpha _s}(\bar \mu )}}{\pi }]^N}\;,
\end{eqnarray}
where
\begin{eqnarray}
{A^{(1)}}({C_1}) = \frac{4}{3},
\end{eqnarray}
\begin{eqnarray}
{A^{(2)}}({C_1}) = \frac{{69}}{9} - \frac{1}{3}{\pi ^2} - \frac{{10}}{{27}}{N_f} + \frac{8}{3}{\beta _1}\ln ({C_1}\frac{1}{2}{e^\gamma }),
\end{eqnarray}
\begin{eqnarray}
{B^{(1)}}({C_1},{C_2}) = \frac{8}{3}\ln (\frac{{{C_1}}}{{2{C_2}}}{e^{\gamma  - 3/4}}),
\end{eqnarray}
\begin{eqnarray}
{B^{(2)}}({C_1},{C_2}) = 2[\frac{{67}}{9} - \frac{1}{3}{\pi ^2} - \frac{{10}}{{27}}{N_f}]\ln (\frac{{{C_1}}}{{2{C_2}}}{e^{\gamma  - 3/4}}) + \frac{8}{3}{\beta _1}\{ {\ln ^2}({C_1}\frac{1}{2}{e^\gamma }) -\nonumber \\ {\ln ^2}({C_2}{e^{3/4}})\}  - \frac{9}{8} + \frac{7}{9}{\pi ^2} + \frac{2}{3}\varsigma (3) + (\frac{5}{{36}} - \frac{2}{{27}}{\pi ^2}){N_f}.
\end{eqnarray}
To achieve more precise results for DY cross section, one can use the NNLO contributions for TMD PDF as was obtained in Sec.2.

The form factor $\widetilde W_{j\bar k}^{}(b_T,Q,{x_1},{x_2})$ in Eq.(\ref{w}) can be  expressed in terms of its perturbative part $\widetilde W_{j\bar k}^{pert}$ and nonperturbative function $\widetilde W_{j\bar k}^{NP}$ as \cite{8}:
\begin{eqnarray}
\widetilde W_{j\bar k}^{}(b_T) = \widetilde W_{j\bar k}^{pert}({b_*})\widetilde W_{j\bar k}^{NP}(b_T).
\end{eqnarray}
with:
\begin{eqnarray}
{b_ * } = \frac{{{b_T}}}{{\sqrt {1 + {b_T}^2/{b_{\max }}^2} }}.
\end{eqnarray}

In our calculations, the cross section of DY process is considered and compared with the available experimental data. Indeed we take the  ${\Lambda _{\overline {MS} }}$ as an unknown parameter that souled be determined by fitting process. We analyze the DY process with BLNY fitting parametrization in the non-perturbative part of the cross section using \cite{6n}:
\begin{eqnarray}
\exp \{ [ - {g_1} - {g_2}\ln (\frac{Q}{{2{Q_0}}}) - {g_1}{g_3}\ln (100{x_1}{x_2})]{b^2}\}, \label{g}
\end{eqnarray}
where ${g_1},{g_2},{g_3}$ are the unknown parameters that should be determined by fitting to the expression in Eq.(\ref{g}) which is corresponding to the square of $(c)$ factor in Eq.(\ref{f}). Using the E288 experimental data (for Q = 5 to 9) \cite{E288} we can obtain the values of  these parameters as well as ${\Lambda _{\overline {MS} }}$ , using CERN subroutine MINUIT \cite{mini}. We calculate the theoretical cross section using b-star method \cite{7}, choosing ${b_{\max }} = 0.5\; Ge{V^{ - 1}} = 0.1$ $fm$ in all part of fitting.

{To get better consistency between the theoretical results  and the related experimental data, the data at each mass bin including  5 $<$ Q $<$ 6, 6 $<$ Q $<$ 7, 7 $<$ Q $<$ 8 and 8 $<$ Q $<$ 9, is multiplied by individual  normalization factors. Then  we have totally  four normalization factors in our calculations.}
\section{Results and discussions}
As mentioned before our purpose is to determine the ${\Lambda _{\overline {MS} }}$ in DY process together with the unknown parameters $g_1$, $g_2$ and $g_3$ at low energy scales, considering transverse momentum distributions by CSS resummation formalism. Therefor we should include the experimental data for which  the transverse momentum distributions have more effect in non-perturbatve part of our calculations. Hence we use the low energy DY data in the region where the transverse momentum  of the lepton pair is much smaller than its invariant mass Q. The reason is that the CSS formalism describes the production of DY pairs better in the central rapidity region \cite{8}. The E288 data from the $p + Cu \to {\mu ^ + }{\mu ^ - } + X$ process at $\sqrt s  = 23.8$ $GeV$ with $x=0.03$ is a good case with those properties. By accessing to the TMDPDFs we are able to calculate the related cross section in DY process.
Using the E288 experimental data, we do the fitting and obtain  the values of the non-perturbative parameters $g_1$, $g_2$ and $g_3$, using the BLNY parametrization  \cite{6n}  and finally ${\Lambda _{\overline {MS} }}$ which is placed in the perturbative part of calculations. The fitting is done by MINUIT programming and the results are:
\begin{eqnarray}
{g_1} = 0.43\pm 0.09\ ,\; {g_2} = 0.67\pm 0.04\ ,\; {g_3} = -0.63\pm 0.07,
\end{eqnarray}
\begin{eqnarray}
{\Lambda _{\overline {MS} }} = 249 MeV \pm 7.
\end{eqnarray}
which is corresponding to ${\alpha _s}(M_z^2) = 0.119 \pm 0.001$ for energy scale of Z-mass boson at the NLO approximation.
{The goodness  fitting with $\chi^2 /{d.o.f.} = 1.12$ and finding for ${\Lambda _{\overline {MS} }}$ a value within the acceptable range indicates that the BLNY parametrization is suitable to describe low energy events.}

We use the obtained numerical results for the unknown parameters and plot the related DY cross section. We then compare the result with the available experimental data and depict them in Fig.2 for $p + Cu \to {\mu ^ + }{\mu ^ - } + X$ reaction at $\sqrt s  = 23.8$ $GeV$ with $x=0.03$.  As can be seen a good consistency is existing between them.

\begin{figure}[htb]
\begin{center}
\vspace{1cm}
\resizebox{0.9\textwidth}{!}{\includegraphics{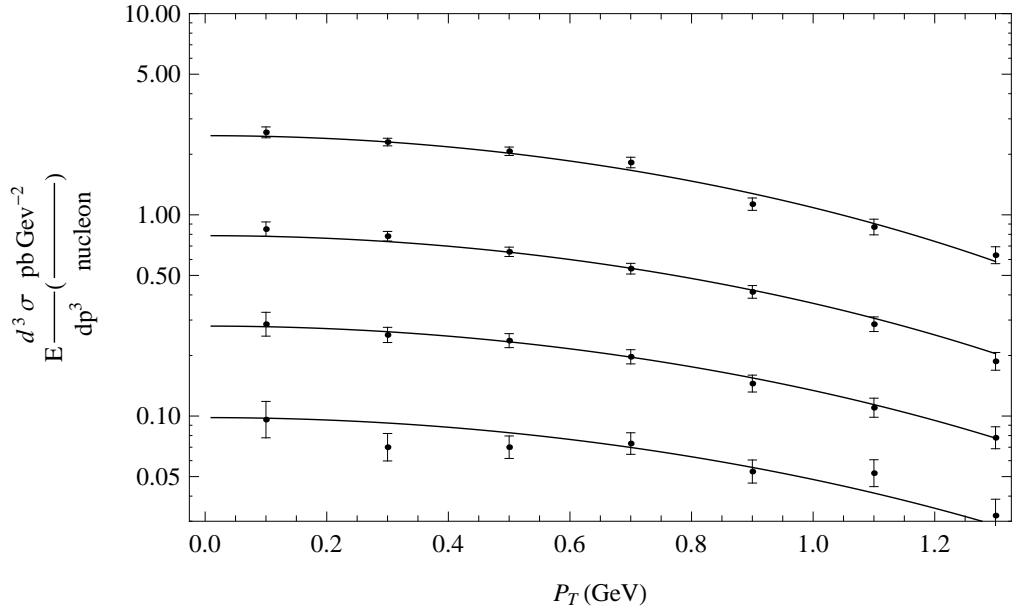}}
\caption{ The theoretical calculation of DY cross section which is compared to E288 experimental data \cite{E288} for the $p + Cu \to {\mu ^ + }{\mu ^ - } + X$ reaction at $\sqrt s$=23.8 GeV.} \label{JR2009PDF}
\end{center}
\end{figure}

\section{Conclusion}
We  considered the evolution of transverse momentum dependent parton distribution function for up quark in CSS formalism up to NNLO approximation which can be easily extended to obtain the TMD PDF for other quarks by similar calculations. In  particular, we have used the TMD evolution kernel at NNLO approximation which to our knowledge, has not been done before. We also indicated that one can directly relate QCD observables to
the underlying dimensional transmutation parameter of the theory, ${\Lambda _{\overline {MS} }}$. We extracted the ${\Lambda _{\overline {MS} }}$ within the appropriate range of perturbative QCD and showed that the TMD formalism is a powerful tool to analyze perturbative and non-perturbative effects in cross section spectra of DY process.  The result of global fit showed that the BLNY parametrization is a good one for the non-perturbative part of cross section and the CSS formalism can formulate and describe properly the E288 experimental data (for low energy Drell-Yan process).

The calculations for DY cross section can be extended to the NNLO approximation which we hope to report them in our further research activity. Employing the experimental data for DY cross section at high energy scales is also a valuable task to increase the precision of our fitting which can be done in future.

\section*{Acknowledgements}
R.T and A.M  acknowledge  Andria Signore which provided us the package of PDF collinear and thanks S.Atashbar Tehrani for useful discussions.

\end{document}